\documentclass[%
 reprint,
 showkeys,
 amsmath,amssymb,
 aps,
]{revtex4-2}

\usepackage{graphicx}
\usepackage{dcolumn}
\usepackage{bm}

\begin{document}

\title{Evolutionary timeline of a modeled cell}

\author{Vrani Ibarra-Junquera}
\affiliation{Laboratorio de Agrobiotecnolog\'ia,\\
Universidad de Colima,\\
Coquimatl\'an, C.P. 28400, Colima, M\'exico.}

\author{Diego Radillo-Ochoa}
\email{Author names are arranged alphabetically. Corresponding author: dradillo1@ucol.mx}
\affiliation{Facultad de Ciencias, Universidad de Colima,
Bernal D\'\i az del Castillo 340, Col. Villas San Sebasti\'an, 
C.P. 28045, Colima, Colima, M\'exico.}

\author{C\'esar A. Terrero-Escalante}
\affiliation{Facultad de Ciencias, Universidad de Colima,
Bernal D\'\i az del Castillo 340, Col. Villas San Sebasti\'an, 
C.P. 28045, Colima, Colima, M\'exico.}

\date{\today}

\begin{abstract}
A theoretical study 
of cell evolution is presented here. 
By using
a toolbox containing
an intracellular catalytic reaction network model
and a mutation-selection process, 
four distinct phases of self-organization were unveiled.
First, the nutrients prevail as the central substrate of the chemical reactions.
Second, the cell becomes a 
\emph{small-world}.
Third,
a highly connected core component emerges,
concurrently with the nutrient carriers becoming the central product of reactions.
Finally, the cell reaches a steady configuration
where the concentrations of the core chemical species 
are described by Zipf's law.
\end{abstract}

\maketitle


\section{\label{sec:intro}Introduction}

A key goal of postgenomic cell biology
is to understand the cellular organization
at its core functional level.
As a result of ongoing efforts in gnome sequencing,
a large number of
highly complex intracellular networks
have been completely mapped out,
revealing common features
not only
among all living organisms,
but also among other natural and even human-made networks.
These findings
lead to the search for the 
fundamental organizational mechanisms
that have
shaped the evolution of life as we currently know it.
A complete introduction to this topic can be found in Ref. \cite{barabasi04},
where most of the technical terminology used in this paper is explained.

Analyzing complex biological networks
on a top-down approach
(from the observational uncovering of the cell structure
to the determination of the specific role of each element)
is a burdensome task,
since the vast diversity and specificity of components,
along with their physical size scale,
makes unveiling the organizational principles difficult.
A key issue here
is that, 
with current imaging techniques,
dynamics inside living cells remain largely inaccessible 
at molecular level.
Even more,
though big steps are being made
in this direction \cite{penedo21},
since these systems 
have been developing over a billion of years of evolution,
this kind of analysis
is constrained by the time at which the networks are sampled,
limiting insight into their evolutionary history.
Efforts are also being done to overcome this time constraint, 
such as
inferring ancestral metabolic networks 
using methods 
to estimate the genetic distance between different bacterial
species 
as given by phylogenetic trees
built following the maximum parsimony criterion \cite{goodman19},
and also
using structure-guided sequence analysis of proteins to explore the patterns of evolution of enzymes responsible for biogeochemical cycles
\cite{bromberg22}.

A fruitful way to address the limitations
mentioned above
is by adopting the philosophy which considers
the cell as the ``minimal unit of life",
i.e.,
a self-sustaining chemical system capable of undergoing Darwinian evolution
\cite{ameta21, benner10}.
This simple cell
can be studied
by means of mathematical models
that describe aspects of the cell
in a bottom-up approach.
The theoretical study of cellular behavior
can be traced back over 50 years to the pioneering works of Kauffman \cite{kauffman69}, 
in which cell complexity is studied from the perspective of random mathematical networks.
His works
belong in the wider context of contemporary network theory, 
which is marked by the discoveries
of surprising similarities
among the structures of a wide variety of real systems
coming from seemingly unrelated areas.
Over the past 20 years,
this branch of biology has experienced an accelerated growth
steered by technological advances 
that have led,
among other advantages, 
to an easier accessibility to higher computational power.
As a result,
several mathematical models 
have been proposed and thoroughly studied,
like
the simple boolean gene-interaction networks \cite{kauffman69.2},
the condensation-cleavage binary polymer model \cite{filisetti14}
or the evolutionary process 
based on an artificial chemistry of catalysed reactions
proposed in
\cite{jain98}.
This kind of simulations
have proved to be useful
for providing insights 
into the underlying principles
of cell evolution.

In this work 
we used a cellular model 
based on catalytic reaction networks, 
which are modified as a result 
of a Darwinian-like evolutionary process of mutation and selection.
A version of the model was first proposed in 1997 by Kaneko and Yomo \cite{kaneko97},
while the corresponding
evolutionary process was introduced in 2006 by Furusawa and Kaneko \cite{furusawa06}.
This model differentiates itself from the rest
because its dynamics are given 
by a system of ordinary differential equations 
that describe the enzymatic kinetics 
of a complex network of unspecified chemical components
which includes passive diffusion of nutrients into the cell
from the environment.
Additionally
it is important to point that,
by including the evolutionary process,
the focus ceases to be on the behavior of a single cell,
and instead the role of the competition among groups of cells is observed,
which ultimately serves as the external input driving the self-optimization process inside a cell.
In previous works it has been shown that different versions of this model are able to reproduce well-known characteristics of living cells,
giving insights into cellular processes such as differentiation \cite{furusawa00}, 
pluripotency \cite{furusawa09} 
and reaction to environmental perturbations \cite{furusawa18}.

The main goal of our work 
is to use the well suited set of tools from graph theory
to provide a basic but comprehensive analysis of the evolution
of the reaction network 
during the mutation-selection process,
as described by a particular version of the Kaneko-Yomo model.
We confirmed that
a self-reproducing system of cells can arise from the simple rules considered in the model and in the evolutionary process.
Beyond that, 
our analysis of the changes 
in the topology of the reaction networks
reveals striking similarities 
with real metabolic networks
and also 
provides new insights into the events 
that could have occurred in the earlier stages
of life development.
Our results are summarized in Fig. \ref{fig:timeline},
which displays
the evolutionary timeline of the modeled cell
as determined by the mutation-selection process.
Motivated by these results,
this manuscript is also meant to serve as groundwork for future research using versions of the model and evolutionary process.
With this aim,
in sections \ref{sec:model} to \ref{sec:mutsel},
we provide a detailed derivation of the expressions 
for the enzymatic kinetics 
and the cellular growth rate,
as well as a description of our computational implementation
of single-cell dynamics and the simulation of the evolutionary process.
This way are filled key voids we have found in the literature about the version of the model we use.
Our results are then addressed
in section \ref{sec:results}.
Finally, 
this paper ends 
in section \ref{sec:conclusions},
with the
presentation and discussion of our conclusions.

\section{Model of the intracellular chemistry}
\label{sec:model}

As mentioned in the Introduction,
our research is enclosed in the conceptual framework 
where a cell is defined
as a self-sustaining chemical system capable of undergoing Darwinian evolution.
As the mathematical model for the internal dynamics of such simple cell,
we use the catalytic reaction network model described in
Ref. \cite{furusawa06}.
In this section we provide 
a detailed derivation of the corresponding
ordinary differential equations,
which seems to be missing elsewhere in the literature.

Let us start by considering the set of variables
$\{n_0, n_1,...,n_N\}$,
which represent the number of molecules of each of the $N$ chemical species that exist within a cell.
In general,
the volume of a cell changes due to variations in intracellular solute content or extracellular osmolality.
Without loss of generality,
in this model we assume that cellular volume is proportional to the total number of molecules inside the cell,
that is,
\begin{equation}
\label{eq:volume}
V=\alpha \sum_i n_i,
\end{equation}
where $\alpha$ is the proportionality constant.
Therefore, 
the intracellular concentration of the species $i$ is given by
\begin{equation}
\label{eq:concentration}
x_i=\frac{n_i}{V},
\end{equation}
such that the internal cellular state can be described by the concentration vector $\mathbf{x}=(x_0,x_1,...x_N)$. 
The rate of change in time $t$ of $x_i$ is then given by 
\begin{eqnarray}
\frac{dx_i}{dt}&=&\frac{1}{V}\frac{dn_i}{dt}-n_i\frac{1}{V^2}\frac{dV}{dt} \nonumber\\
&=&\frac{1}{V}\frac{dn_i}{dt}-x_i\frac{1}{V}\frac{dV}{dt}.
\end{eqnarray}
Let us define
\begin{eqnarray}
R_i&\equiv& \frac{1}{V}\frac{dn_i}{dt}\, ,\\
C_i &\equiv& -x_i\frac{1}{V}\frac{dV}{dt}\, ,
\end{eqnarray}
such that
\begin{equation}
\frac{dx_i}{dt}=R_i+C_i.
\end{equation}
Note that $R_i$ is related to the change in the amount of molecules of the species $i$, 
whereas $C_i$ relates to the overall change in cellular volume.

In this model, the number of molecules of each species as a function of time
is determined by the dynamics of a catalytic reaction network.
Reaction kinetics are assumed to be described by the law of mass action in steady state. 
For simplicity,
the only interaction between the environment and the cell considered in this model is
the passive diffusion of
a few species through the cell membrane.
From now on these permeable species are referred as \emph{nutrients},
and their diffusion is aided by specific \emph{carrier} species.
With these considerations, $R_i$ is defined as
\begin{multline}
R_i (\mathbf{x}) \equiv \sum_{j,\hspace{1pt}l} \sigma_{j, \hspace{1pt} i, \hspace{0.8pt} l} \hspace{3pt} x_j x_l - \sum_{j',\hspace{1pt}l'} \sigma_{i, \hspace{2pt} j', \hspace{2pt} l'} \hspace{3pt} x_{i} x_{l'} \\
\Big( + \hspace{3pt} Dx_{m_i}(\textsc{x}_i-x_i) \Big), 
\end{multline}
where the first two terms represent the enzymatic kinetics involving the species $i$ 
as product or substrate, 
with each reaction correspondingly catalyzed by the species $l$ and $l'$. 
The factor $\sigma_{i,j,l}$ is equal to $1$ 
if reaction $i+l\longrightarrow j+l$ takes place, 
and $0$ otherwise. 
The third term is written in parentheses 
to denote that it is only added to the equations of the nutrient species
to include the corresponding diffusion process
with coefficient $D$ and its respective carrier $x_{m_i}$.
It is worth emphasizing that considering the role of a carrier species
results in an effective time dependent diffusion coefficient,
which allows the cell to self-regulate its exchange of nutrients with the environment.

Next,
in order to express $C_i$ in terms of the vector $\mathbf{x}$, 
let us consider the derivative of Eq. (\ref{eq:volume}),
\begin{equation}
\label{volume_derivative}
\frac{dV}{dt}=\alpha \sum_j \frac{dn_j}{dt}.
\end{equation}
Multiplying both sides by $-x_i/V$
and substituting the definition of $R_i$ and $C_i$, 
we obtain
\begin{equation}
C_i(\mathbf{x}) = -\alpha x_i \sum_j R_j.
\end{equation}
Finally, 
the rate of change for the intracellular concentration of a species $i$ can be expressed as
\begin{equation}
\label{eq:model_der}
\frac{dx_i}{dt} = R_i - \alpha \hspace{2pt}x_i \sum_{j}R_j \hspace{3pt}.
\end{equation}
In the case $\alpha=1$, 
this equation corresponds to the one presented in Ref. \cite{furusawa06}.

Typically, 
a large number $N$ of species is considered in the model,
therefore the solutions to the above system must be obtained numerically.
We do this by approximating $R_i$ using an implementation of the fourth order Runge-Kutta method for many variables.
Although sophisticated multi-step methods offer higher accuracy of the numerical approximation,  
with reasonable increase in the computational requirements,
we have chosen the relatively simple Runge-Kutta method, 
taking into account that Eq. (\ref{eq:concentration}) 
provides a convenient test for accuracy of our numerical results;
identity
\begin{equation}
\label{eq:norm}
\sum_i x_i = \frac{1}{\alpha},
\end{equation}
must always be satisfied
during the whole cell evolution.
The value of $\alpha$ is established by the selection of the initial conditions for the concentrations.

As mentioned before,
the dynamics of the concentrations of the different chemical species are dictated by a catalytic reaction network.
Let us now describe
the mathematical representation we used to model it. 
The reaction network can be represented by a \emph{directed graph},
a mathematical object consisting of a set of \emph{nodes} connected by \emph{edges} that have a definite direction.
Each node in the network corresponds to a different chemical species.
A pair of nodes is connected by an edge if there is a reaction 
that involves its respective species either as subtract or product. 
The edge will be directed towards the node corresponding to the product of the reaction.
Every edge possesses an attribute associated with the catalyst of the reaction it represents.
Schematically, a reaction in this network could be represented as
\begin{equation*}
    i \xrightarrow{\hspace{15pt} l \hspace{15pt}} j.
\end{equation*}
The structure of this graph additionally complies with a number of
biologically inspired
constraints.
As we will see in the following sections, 
cell growth 
is central to our study,
considering it to be due to biosynthesis processes,
reactions on the network are set to occur on a single direction;
such that there are no symmetric pairs of directed edges,
restricting its structure to one of an \emph{oriented graph}.
Moreover, 
the nutrient species are not allowed to be product of any intracellular reaction,
so that they only grow by permeating to the cell from the environment.
Also, 
since facilitated diffusion typically involves the transport of ions and polar molecules such as carbohydrates and amino acids,
nutrients 
are not used in this model as catalyst enzymes for any reaction.
Neither are nutrient carriers,
due to their specific role.
With these constraints,
the catalyst for each reaction is chosen randomly from the set of all available catalysts, 
verifying that
the reactions satisfy a one-to-one correspondence from the substrate-catalyst pair $i+l$ to the product-catalyst pair $j+l$, where $l \neq i,j$.

Let us end this section by recalling that 
the binary fission time of a cell
strongly depends on the organism 
and even in the given tissue of a multicellular organism.
It varies over a broad range spanning from minutes to years.
Taking this into account,
in Eqs. (\ref{eq:model_der}) 
we are using dimensionless time.    

\section{Cell growth rate}
\label{sec:growth_rate}

Cellular growth rate plays a central role in this work.
Following Ref. \cite{furusawa06}, 
it is defined as the inverse of the duplication time $t_d$, 
i.e., 
the time it takes for a cell to double its initial volume.
Recall that in this model the volume is proportional to the number of molecules inside the cell (see Eq. (\ref{eq:volume})).
However, the dynamic variables used in
Eqs. (\ref{eq:model_der}) are the concentrations of the different chemical species within the cell,
$\mathbf{x}$,
which are in turn determined by the cellular volume, 
as seen in Eq. (\ref{eq:concentration}).
As a consequence,
any attempt to combine these equations to obtain an expression for the volume in terms of the concentrations results in an inconsistent system.
Therefore,
to obtain an expression to determine $t_d$ becomes non trivial.
The expression to calculate the growth rate
we use here
is different from the one given in Ref. \cite{furusawa98}.
and the related literature.
To derive it,
let us start by rewriting Eq. (\ref{volume_derivative}) as follows,
\begin{equation}
\frac{dV}{V}=\alpha \sum_j R_j (\mathbf{x}) \hspace{2pt} dt.
\end{equation}
This expression can be integrated over a time interval $t_0$ to $t$,
\begin{equation}
\label{volume ratio}
\ln \frac{V_t}{V_0} = \alpha \int_{t_0}^{t} \sum_j R_j (\mathbf{x}) \hspace{2pt} dt.
\end{equation}
where $V_0$ is the initial cellular volume and $V_t$ is its value at time $t$.
Since at $t=t_d$ the volume is $V_t=2V_0$,  
the condition for the duplication time becomes,
\begin{equation}
\label{eq:duplication time}
\alpha \int_{t_0}^{t=t_d}\sum_j R_j (\mathbf{x}) \hspace{2pt} dt\hspace{3pt}=\ln 2.
\end{equation}
In our work 
$R_j (\mathbf{x})$ is calculated from the numerical approximation for the temporal evolution of $\mathbf{x}$, as mentioned in Sec. \ref{sec:model}.
This means that the definite integral in Eq. (\ref{eq:duplication time}) must be numerically approximated as well. 
In our implementation, this is done  using the composite Simpson's rule, 
where an approximation $S$ to the value of an integral is calculated using $s+1$ equally spaced points with a step size $h$ over a given time interval $(t_i,t_f)$ as
\begin{equation}
\begin{split}
\int_{t_i}^{t_f} \sum_j R_j (\mathbf{x}) \hspace{2pt} dt \approx S(t_i,t_f) \\
= \frac{h}{3}\Big(f_0+4f_1+\sum_{j=1}^{s/2-1} \big[ \hspace{2pt} 2f_{2j}+4f_{2j+1} \hspace{2pt} \big] + f_{s} \Big).
\end{split}
\end{equation}
Here $f_n$ corresponds to the value of $\sum_j R_j(\mathbf{x}_n)$ at the time $t_n=t_i+n\hspace{1pt}h$.

Nevertheless,
our task is not to approximate the integral over a given interval,
but to find the upper limit satisfying 
duplication condition (\ref{eq:duplication time}).
Furthermore,
as we will describe in Sec. \ref{sec:mutsel}, 
this have to be repeated millions of times in a single simulation.
Considering this,
we implemented an efficient algorithm to estimate the value  of $t_d$.
Let $S_T$ be an approximation to the value of the definite integral in Eq. (\ref{volume ratio}), 
calculated by summing a number of approximations $S$ over consecutive sub-intervals within $t_0$ to $t$. 
We start by calculating sequential approximations $S$ 
using a broad integration range and time step 
until the cellular volume exceeds its duplication value
\begin{equation}
   S_T > \ln 2.
\end{equation}
Once in the vicinity of $t_d$, 
a couple of refining stages repeatedly cut in half both
the integration range and time step 
until (\ref{eq:duplication time}) is satisfied within a certain tolerance $\epsilon$ defined as
\begin{equation}
\label{eq:tolerance condition}
\left| \alpha \int_{t_0}^{t=t_d}\sum_j R_j (\mathbf{x}) \hspace{2pt} dt-ln\hspace{3pt} 2 \right| < \epsilon.
\end{equation}

\section{The evolutionary process}
\label{sec:mutsel}

The \emph{mutation-selection} process is
a simple stochastic mechanism
applied to groups of cells
that results in successive modifications of the 
corresponding reaction networks
in favor of faster cellular growth.
During this process,
the reaction network of a cell can be modified
by adding a new edge between two randomly selected nodes.
This action is called a \emph{mutation}.

It is important to note that
we distinguish between the time evolution of the chemical species inside the cell, as described in Sec. \ref{sec:model},
and the changes on the structure of the internal network of the cells
as they evolve. 
As already mentioned, 
in real life,
the evolution of intracellular chemical species until binary fission
takes from minutes to thousands of hours.
On the other hand, 
a mutation
could take several thousands of years
to prevail.
With such dramatic time differences 
it is impossible to use real time to trace the evolution of cellular networks.
Therefore,
we use the approach described in Ref. \cite{barabasi16}
and our \emph{event time}
will be given by changes in the network topology 
through the mutation-selection process,
as described below.

This process begins with $m=2N$ initial cells 
that share an identical internal state $\mathbf{x_0}$,
but
whose graphs are generated by different mutations to a given initial network.
These $m$ cells are allowed to evolve in time, 
and their growth rate
is estimated as described in Sec. \ref{sec:growth_rate}.
Then, 
the $n$ cells with the highest growth rate 
are selected to be the \emph{mothers} cells of the first \emph{generation}.
Each generation is conformed by 
a set of $n \times m $ \emph{daughter} cells. 
The network of a given daughter cell is generated 
as a mutation
of one of the mothers.
This is done $m$ times for each of the $n$ mother cells.
Consistently with the fact that a daughter cell is not an exact replica of its mother,
the daughters inherit
as initial conditions
the internal state of its mother at the duplication time
plus a tiny perturbation on the values of the chemical concentrations.
Subsequently,
daughter cells are allowed to evolve, 
and growth rate is estimated.
The $n$ ones growing faster up to division
are selected to be the mothers of the next generation, 
and so on.

\begin{figure*}[!htbp]
\includegraphics[scale=1]{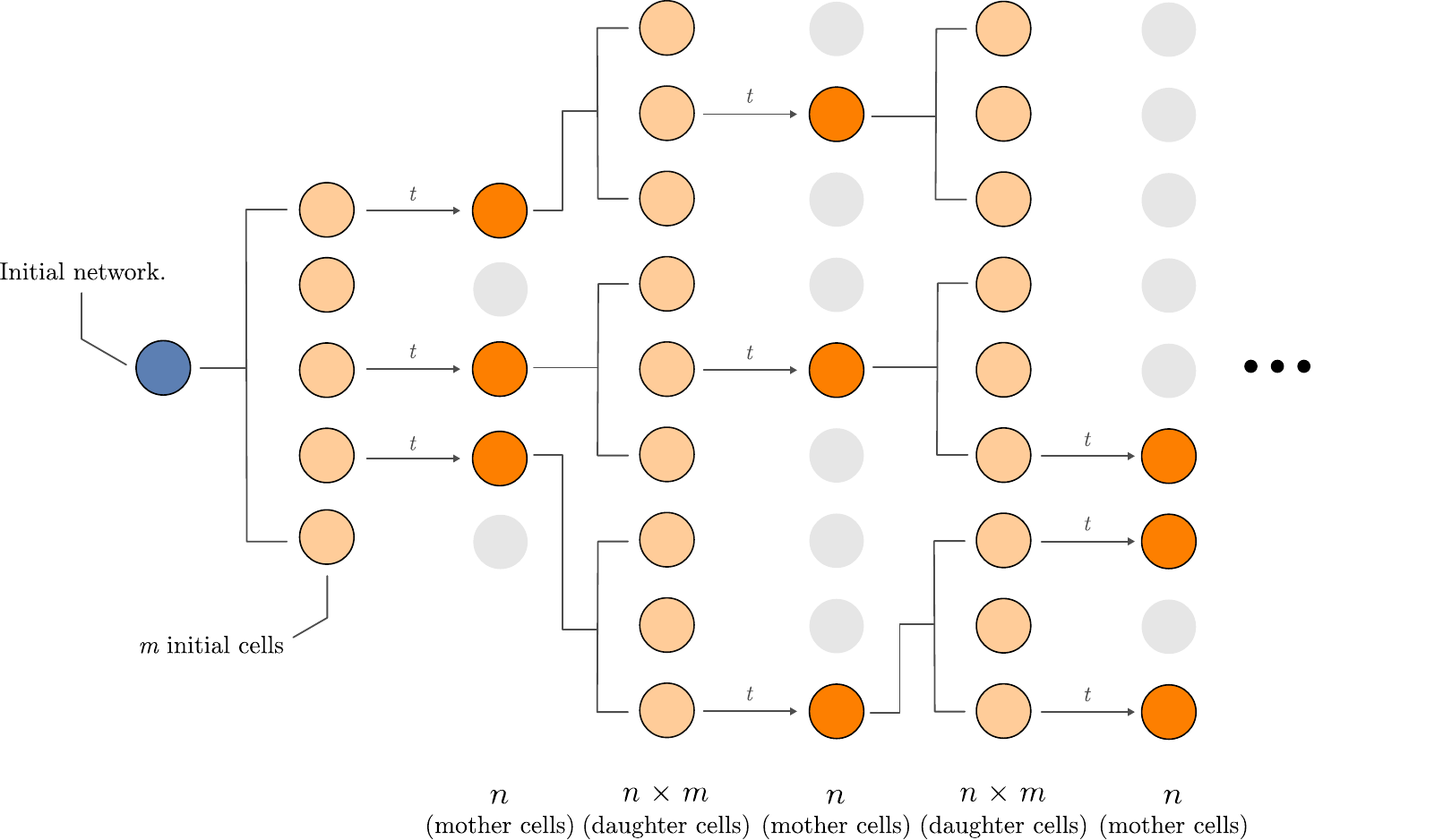}
\centering
\caption[Mutation-selection process diagram.]{Mutation-selection process diagram. Columns of light orange circles represent the set of $n \times m$ daughter cells of each generation.}
\label{fig:mutsel}
\end{figure*}

This process is illustrated in Fig. \ref{fig:mutsel}.

\begin{figure}[!htbp]
\includegraphics[scale=1]{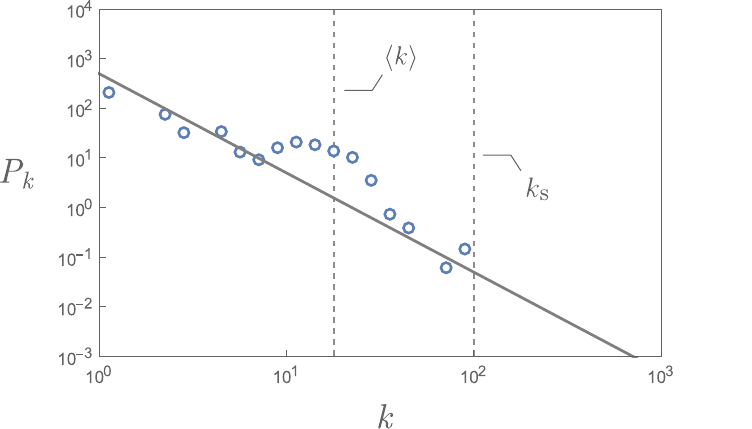}
\centering
\caption[Degree distribution in the case of high-degree saturation.]{Characteristic peak in the degree distribution in the spurious case of high-degree saturation. Here we show a network with $N=500$ and $\langle k_{initial} \rangle = 1$ at the generation $g=5000$. The saturation threshold was set at $k_{S}=100$. The solid line represent a power-law with exponent $-2$.}
\label{fig:degreesat}
\end{figure}

Let us end this section by emphasizing the underlying computational difficulties
involved in our study.
As mentioned before,
a large number $N$ of chemical species must be considered
and
we follow the mutation-selection process for at least $4N$ generations,
each of which consist of a set of $n \times m$ cells.
For a typical simulation
this requires to numerically solve 
millions of systems of nonlinear coupled ordinary differential equations including hundreds of variables.
Besides the obvious requirement of the parallelization of the computation,
it was also necessary to optimize the different algorithms
as well as 
design an efficient flow structure for the whole simulation.
Nevertheless,
this efficiency can cause spurious results
that must be brought under our control.
For instance,
for a given set of $N$ nodes,
the maximum possible 
\emph{node degree}
(number of paths linked to a node)
is $k_S= N-1$, 
i.e., when it is connected to every other node.
This way,
for a node with degree $k$,
there remain 
$N-1-k$ possible mutations to be added.
As the degree approaches $k_S$,
the probability of the corresponding node
being chosen from the mutations list 
is significantly reduced.
We call this effect \emph{high-degree saturation}.
Fortunately,
in most cases 
this saturation threshold $k_S$ is high enough
to cause no concern.
Our implementation takes advantage of this
and effectively reduces the execution time from $O(n^2)$ to $O(n)$
by lowering the saturation threshold to $(N-1)/2$ or less.
This is done by restricting the sample space of the possible mutations
to the links of a complete 
(or near complete) 
random oriented graph,
which is generated as a template
prior to the execution of the simulation.
Nevertheless,
as saturation happens
it can be easily identified because
the degrees statistics
will follow the Gaussian distribution 
of the corresponding template graph.
This will be reflected  
as a characteristic peak centered at the average node degree 
$\langle k \rangle = \langle k_{initial} \rangle + g/N$, 
where $g$ is the current generation
(see Fig. \ref{fig:degreesat}).
Therefore,
in our implementation,
we are dictating the asymptotic outcome
of our mutation-selection process,
though not the way it is reached.
For all the simulations
that we discuss in the following section,
we follow closely the maximum degree of the nodes
reached after each generation 
to make sure it is sufficiently under the saturation threshold.
This gives us a number of reliable generations for a given simulation.
Furthermore,
it is worth emphasizing that
for each simulation we generate a different template graph.
 
\section{Results}
\label{sec:results}

The results to be presented in this section
were obtained using 
four nutrients,
the external nutrient concentration $X_i=0.2$,
the diffusion coefficient $D=4$,
and the number of mother and daughter cells for each generation
to be $n=10$ and $m=2N$, correspondingly
(where $N$ is the number of species).
These values are borrowed from Ref.\cite{furusawa06},
where it is claimed that their results are robust 
with respect to variations of the model and simulation parameters.
Furthermore, 
we choose the initial values for the concentrations randomly,
but such that $\alpha=1$ in Eq. (\ref{eq:norm}),
with random perturbations of the order of $10^{-6}$ 
for the concentrations inherited by the daughters.
Finally,
we set $\epsilon = 10^{-5}$ in 
Eq. (\ref{eq:tolerance condition}) 
for the tolerance in the estimation of the cell duplication time.

As a first result it is convenient to mention
that most of our outcomes were found to be independent of the number of chemical species,
for $N$ sufficiently large.
Taking this into account,
we often generalize their display by labelling them using 
the number of nodes $N$,  
which in our simulations took the values $250, 500$ and $1000$.

\subsection{Degree of connectivity}

As initial configuration
for our simulations, 
we used
random networks in three different connectivity regimes 
around the critical connectivity point.
At this point 
a network has an average degree of $\langle k \rangle = 1$,
i.e., 
the average number of edges per nodes is one.
In the sub-critical regime
($\langle k \rangle < 1$)
consist of small sub-networks isolated from each other,
while
in the super-critical regime
($\langle k \rangle > 1$) 
are characterized by the existence of a dominant large-sized sub-network known as \emph{giant strongly connected component} (\textsc{gscc}).
Besides $\langle k \rangle = 1$,
we also consider graphs with an initial average degree 
of $\langle k \rangle = 0.5$
and $\langle k \rangle = 4$. 
Note that in all three cases, 
the connectivity stays below the threshold
for
fully connected 
random directed graphs, 
$\langle k \rangle = \ln N$
\cite{palasti66}.

\begin{figure*}[!htbp]
\includegraphics[scale=1]{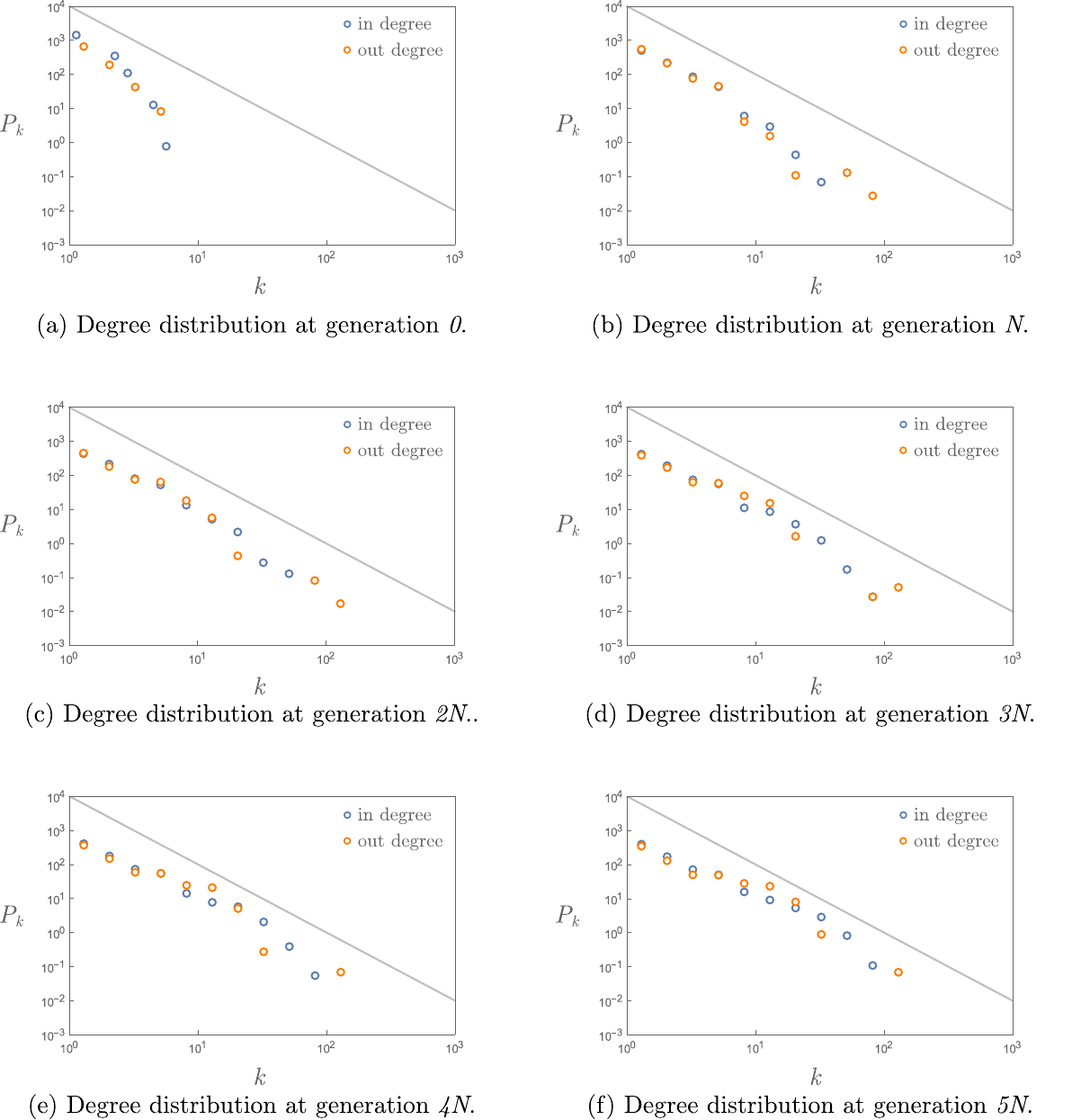}
\centering
\caption[Evolution of the degree distributions.]{Evolution of the in- and out-degree distributions. Here we show the results from a $1000$ nodes network with $\langle k_{initial} \rangle = 1$. The solid line represents a power-law with exponent $-2$.}
\label{fig:degree_evo}
\end{figure*}

\begin{figure}[!htbp]
\includegraphics[scale=1]{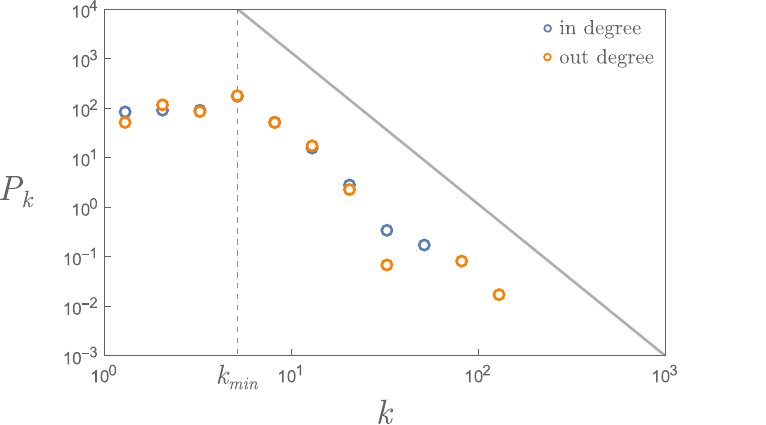}
\centering
\caption[Degree distribution in the case of high initial connectivity.]{Degree distribution of a network at generation $2N$ starting off with a configuration with $\langle k \rangle = 4$. The solid line represents a power-law with exponent $-3$. The dashed line indicates the value of $k_{min}$. Here we show the results for a network with $1000$ nodes.}
\label{fig:degreedist_k4}
\end{figure}

Since we start with random networks,
at the beginning of the simulations
both, 
the in- and out-degree distributions, 
obey binomial laws,
as observed in Fig. \ref{fig:degree_evo}(a)
\footnote{Plots were built using logarithmic binning to correct for the sampling bias introduced by a typical lineal binning  
(see Ref. \cite{barabasi16}, \S 4.12 \emph{Advanced Topic 3.B: Plotting Power-laws}).}.
Afterwards, 
we found that the behavior of the topology of the graphs
during the mutation-selection process
strongly depends on the initial connectivity of the networks.
In general, 
it is observed that by the $N$-th generation
the degree distribution already shows a well defined pattern. 
In the cases starting with a low connectivity
($\langle k \rangle = 0.5, 1.0$),
the degree distribution 
is described by a power-law with an exponent near $-2$,
as seen in Fig. \ref{fig:degree_evo}(b).
This value signals the emergence of 
a hub-and-spoke network 
where the hubs are in contact with a large fraction
of all nodes
\cite{barabasi04}.
For these cases,
as the simulation continues, 
the size and number of hubs increase, 
extending the distribution toward higher connectivities. 
The functional form of the distribution is maintained during the rest of the mutation-selection process 
as seen in Figs. \ref{fig:degree_evo}(c)-\ref{fig:degree_evo}(f).
On the other hand,
the degree distribution developed
in the cases that started with a high connectivity 
($\langle k \rangle = 4$)
deviate from a pure power-law,
showing a low-degree cutoff below a certain $k_{min}$,
as observed in Fig. \ref{fig:degreedist_k4}. 
This behavior
is not shown in analogous simulations
(using $\langle k \rangle = 4$)
previously conducted by Furusawa and Kaneko \cite{furusawa06}.
Yet,
these authors report a power-law distribution with an exponent near $-3$,
which in turns
agrees with a fit restricted to the scaling region observed in our simulations.
As explained in Ref. \cite{barabasi04},
in the case of a pure power-law with exponent $-3$,
hubs with the largest degree of connectivity 
are linked to a small fraction of all nodes,
which is to be contrasted with the case with exponent $-2$
described above.
Noting that $\langle k \rangle = 4$
is rather close to $\ln N$,
the full connectivity threshold for a random
directed graph, 
we performed one more simulation with an intermediate initial connectivity of $\langle k \rangle = 2$.
We found its behavior to be qualitatively the same as the one obtained for low initial connectivity, 
with a degree distribution that follows a power-law with an exponent near $-2$.

Our results suggest that the evolution observed for 
$\langle k_{intial} \rangle = 4$ 
is not distinctive of the mutation-selection process,
but rather an artifact of a particular setup 
that imposes constraints on the graph topology.
Since nodes can only gain connections and
there is a fixed number of nodes,
a random network with a high initial connectivity 
directly limits the amount of possible low-degree nodes during a simulation,
leading to the observed cutoff in the low connectivity region.
Even if this behavior is consistent with a wider definition of the scale-free property, 
the excellent fit 
to a pure power-law 
of the degree distribution of real metabolic networks \cite{broido18}
should discourage the use of highly connected initial states.
Even more,
if we identify connectivity with complexity,
it makes more sense,
from an evolutionary perspective,
to start from simple cells
and evolve into more complex ones.

\subsection{Small world structure}
\label{sec:<d>}

One universal characteristic of complex networks 
is that they usually describe
\emph{small-worlds}.
In these networks the distance between any pair of nodes,
no matter how disconnected they seem to be,
is usually considerably shorter than naively expected.
In graph theory,
the distance between two nodes
is measured as the number of edges
in the shortest path that connects them.
The \emph{average minimum distance}
among all pairs of nodes,
$\langle d \rangle$,
characterizes the overall network interconnectivity,
and in general
it quantifies the flux or transmission efficiency of the system.
In the context of a metabolic network,
$\langle d \rangle$ indicates the average length of the shortest pathway 
between every pair of metabolites.
In a small-world 
it is known that,
for graphs with similar degree of connectivity,
the average minimum distance satisfies
$\langle d \rangle \sim \ln N$.
This implies that 
$\langle d \rangle$ depends slightly
on network size, 
a characteristic typically observed in the structure of 
metabolic networks \cite{jeong00}.

\begin{figure}[!htbp]
\includegraphics[scale=1]{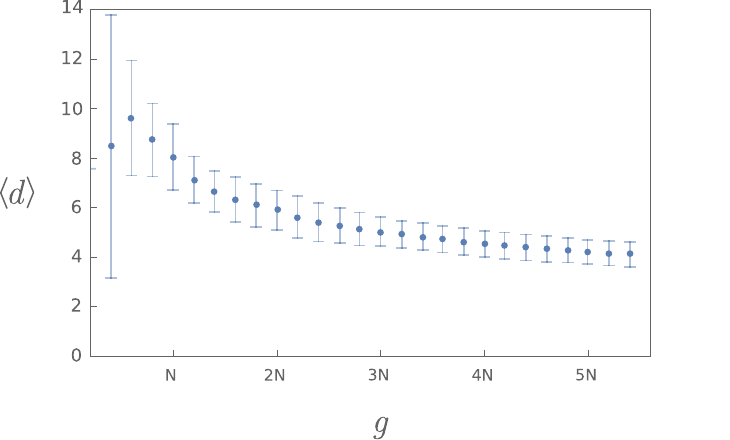}
\centering
\caption[Evolution of the average minimum distance.]{Evolution of the mean and standard deviation of the average minimum distance ($\langle d \rangle$) during the mutation-selection process of different simulations with network sizes varying from 250 nodes up to 1000 nodes. The initial configuration for all simulations is a random network with 
$\langle k \rangle=1$.}
\label{fig:minimum_distance}
\end{figure}

When studying the connectivity properties of sparse networks, 
as our modeled cells,
it is useful to describe exclusively the \textsc{gscc},
as it reflects the connectivity of the whole network
\cite{hong03}.
Taking into account the results in the previous section, 
from now on
we consider simulations that 
started with a connectivity $\langle k \rangle=1$.
Therefore,
as already mentioned,
after a few mutations
there should arise a \textsc{gscc}.
We verified that this indeed happened in our simulations.
Taking this into account,
to study the behavior of $\langle d \rangle$
we follow the evolution of the \textsc{gscc} of three networks
with $N=250, 500$ and $1000$ nodes, 
respectively.
Then we determine the mean and standard deviation
of $\langle d \rangle$ for these cases.
Even if this sample of three cases is poor,
Fig. \ref{fig:minimum_distance}
allows us to conclude
that the mutation-selection process leads to small-worlds,
because it is observed
that the standard deviation decreases drastically
with the pass of the generations,
while
the average minimum distance converges 
to a value around $4$ independently of the network size
in an comparable number of generations 
(parameterized in the figure by $N$).
Notice
that the asymptotic value for $\langle d \rangle$
is remarkably close to those from real metabolic networks,
which fall in the interval 
from 2 to 5, 
depending on the organism \cite{jeong00}.

\subsection{Central species}

While the average minimum distance provides a large-scale characterization of the network, 
\emph{centrality} measures provide information about specific 
nodes that play an important role in the overall network dynamics.
Here we are going to focus on the \emph{closeness centrality}, which for a node is defined as the inverse of its average minimum distance from every other node.
In the case of directed networks the closeness centrality can be measured both inwardly and outwardly from the nodes.
Through this parameter it is possible to objectively identify the  central metabolites on a metabolic network \cite{fell01}.
Centrality analysis helps to understand the metabolic system
not only at a structural level,
but at the base of its functional organization,
providing additional clues about the mechanisms
that shape the networks
as they evolve.

\begin{figure*}[!htbp]
\includegraphics[scale=1]{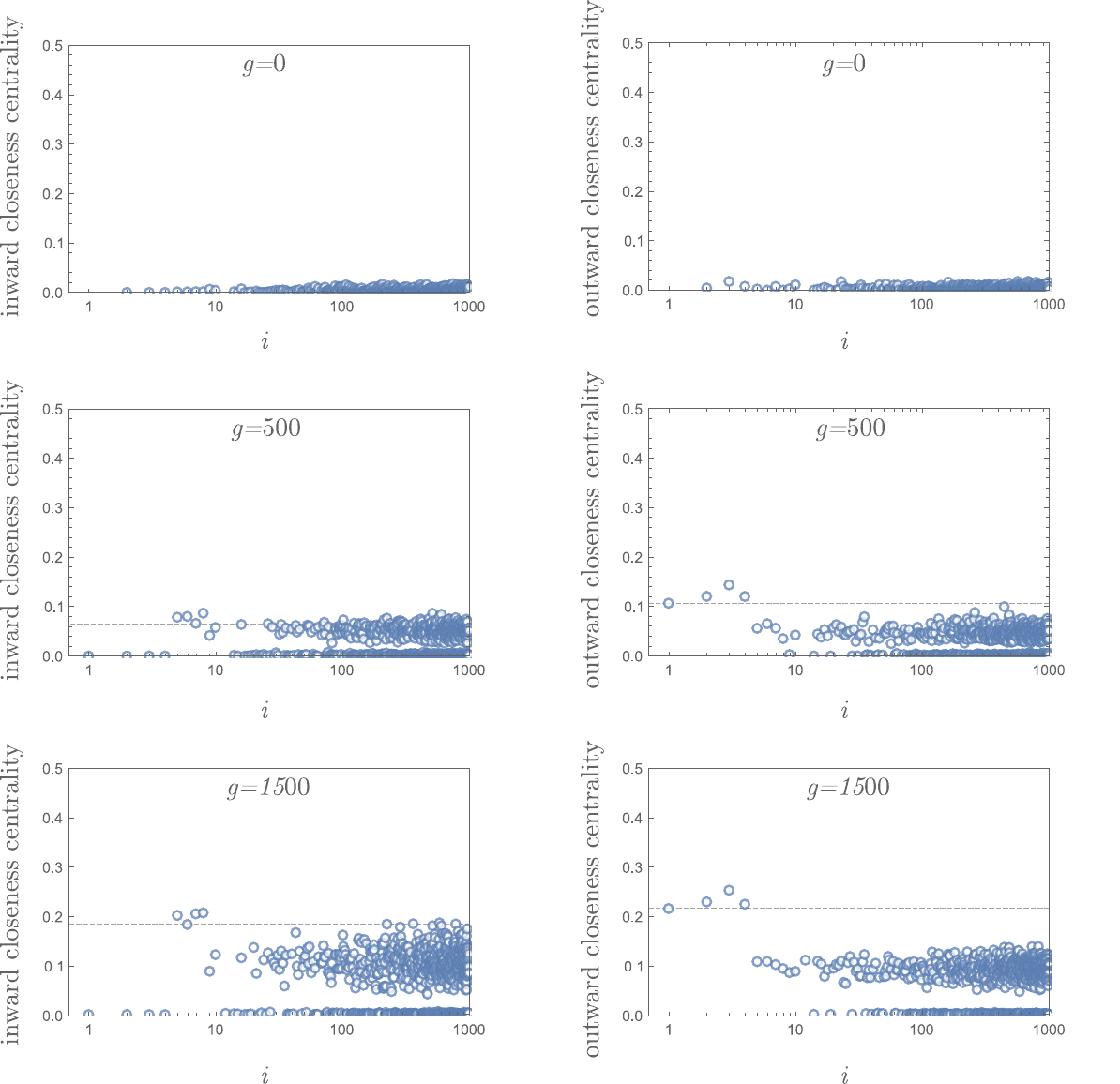}
\centering
\caption[Evolution of the centrality of the nodes.]
{Evolution of the 
inward (left column)
and
outward (right column)
centrality of the nodes during the mutation-selection process with a 1000 nodes network. The first four nodes correspond to nutrient species and the following four are their respective carriers.}
\label{fig:centrality_evo}
\end{figure*}

As shown in the right panels 
of Fig. \ref{fig:centrality_evo},
in our simulations
nodes corresponding to nutrient species
developed the highest outward closeness centrality, followed by randomly distributed nodes.
This result can be explained by the fact that,
if by chance
any other non-nutrient species 
gets involved as substrate in a large number of reactions,
these reactions will stop after the given species is exhausted.
In contrast,
if the inner nutrients concentrations decline,
a gradient with respect to the external nutrients concentrations
builds up,
which
leads to an increase in nutrient intake from the environment.

On the other hand,
in the left panels of Fig. \ref{fig:centrality_evo},
we observe that
the carrier species developed the highest inward closeness centrality.
This seems to happen several generations
after the nutrients dominated the outward centrality,
implying that,
at this point of the evolution,
when the connectivity of the graph has grown significantly, 
the above mentioned induced gradient
seems to no longer be sufficient
to sustain the nutrient intake from the environment 
needed for the reactions.
This problem is henceforth solved for those cells
with networks able to yield
higher concentrations of nutrient carriers.

Within the limitations of the catalytic reactions model
stated in Sec. \ref{sec:model},
these results are compatible with what
is known for the metabolic networks of living organisms
\cite{hong03}.

\subsection{Hierarchical structure}

A wide variety of real systems present networks that can be subdivided into \emph{modules},
weakly interacting subsets of strongly interconnected nodes. 
Modules had attracted a lot of interest because they are typically associated with specific system functions. 
The development and optimization of algorithms for module detection in large networks is currently one of the main challenges for graph theory, 
due to the extensive computational requirements and the selection of the search criteria. 
Some examples of these search criteria are 
the modularity maximization 
and the suppression of high centrality links, 
which respectively correspond to the Greedy and Girvan-Newman algorithms \cite{clauset04}.

The existence of a hierarchical structure in networks from both natural and artificial systems is a fact \cite{corominas13}. 
These structures consist of modules constituted by smaller sub-modules in a recursive fashion.
This property is reflected in the distribution of the 
\emph{clustering coefficient} of the network nodes
behaving as $C_k \sim k^{-1}$. 
Taking this into account, 
it is possible to detect quantitatively 
the presence of a hierarchical structure on a network \cite{ravasz03}. 
In particular, this is the behavior observed in metabolic networks \cite{barabasi04}.

\begin{figure}[!htbp]
\includegraphics[scale=1]{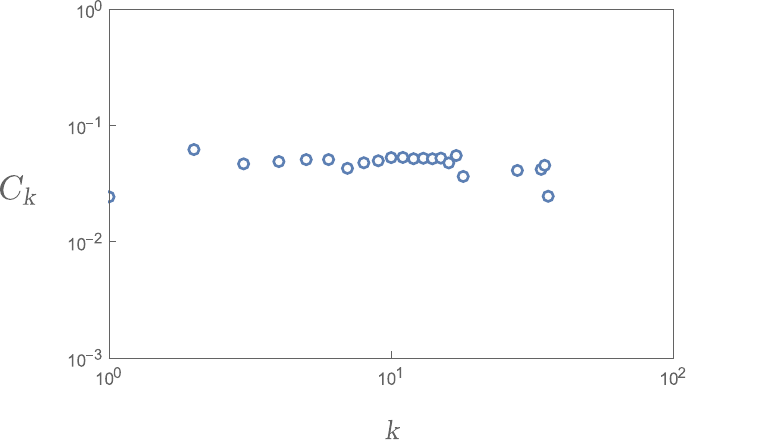}
\centering
\caption[Typical clustering distribution of the networks developed during the simulations.]
{Typical clustering distribution of the networks during the simulations. 
Here we show the distribution of a $250$ nodes network after $1250$ generations.}
\label{fig:clustering_dist}
\end{figure}

Our analysis of the networks resulting from the mutation-selection process 
reveals a different behavior; 
$C_k$ is mainly independent of the node degree, as observed in Fig. \ref{fig:clustering_dist}. 
At an early stage of the evolutionary process, 
this behavior is expected, 
because of the properties of the Erdös-Reyni random network model. 
Nevertheless, 
we found that the mutation-selection mechanism have little effect on this behavior.
To some extent, 
this is also to be expected owing to the similarity of the mechanism considered here
to the more general Albert-Barabási preferential attachment model, 
where the clustering distribution is also independent of the node degree 
\cite[\S 9.3 \emph{Hierarchical Clustering}]{barabasi16}.

\subsection{Intracellular species concentration}

The results presented in the previous subsections
give us a good idea of how the topology of the graph
representing a cell
varies 
as dictated by the mutation-selection process.
Topology
determines the general pattern of
a cell metabolism,
but even for a fixed graph,
the particular realization of the evolution of the internal cellular state
depends on the initial conditions for 
Eqs. (\ref{eq:model_der}).
Then, it is interesting to determine what is universal in the
intracellular dynamics
as it changes during the evolutionary process.

\begin{figure*}[!htbp]
\includegraphics[scale=1]{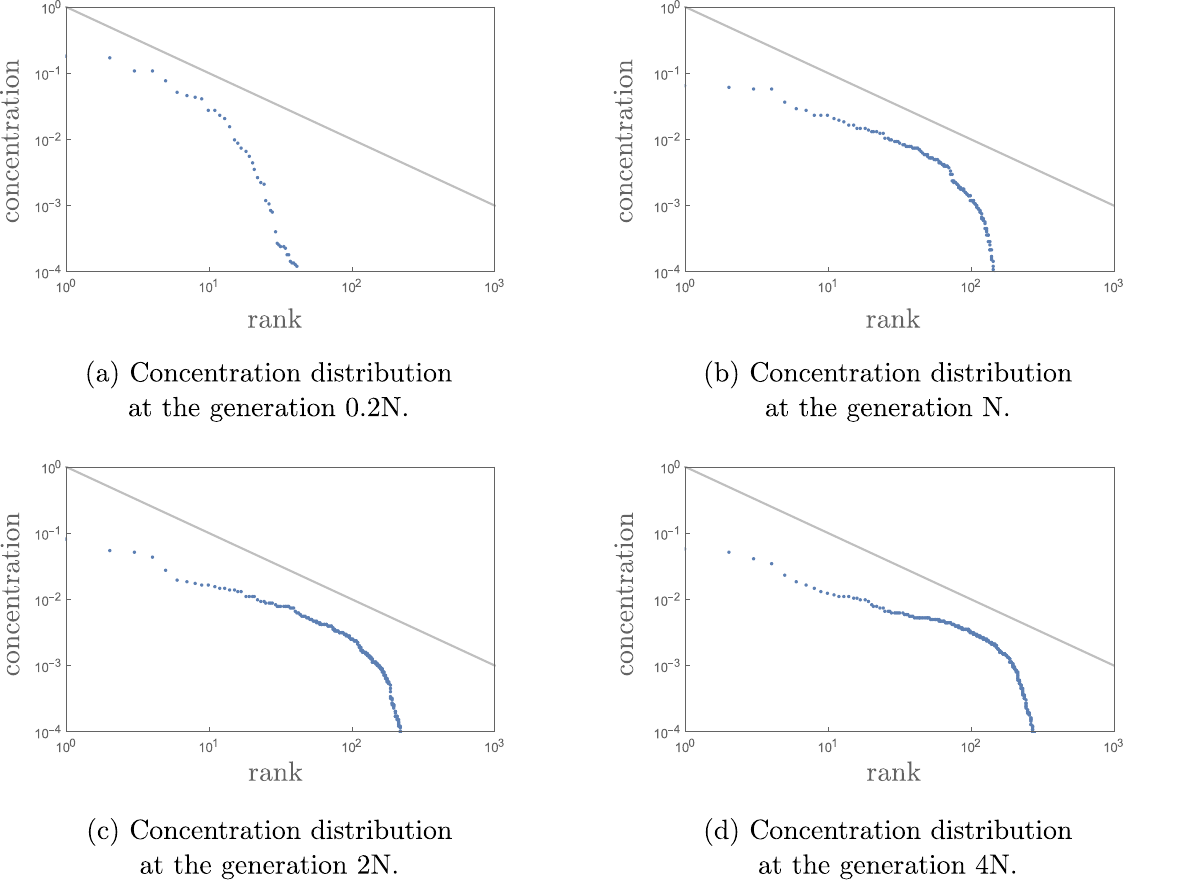}
\centering
\caption[Evolution of the intracelullar chemical concentration distribution.]
{Evolution of the intracelullar chemical concentration distribution over the course of a simulation. 
This distribution is built after ranking the species from highest concentration to lowest concentration.
The solid line corresponds to a power-law with exponent $-1$.
Here we show the result for a network with $N=500$ nodes.}
\label{fig:concentration_evo}
\end{figure*}

In all considered cases, 
starting from a random internal state,
after a relatively small number of generations,
a scaling region arises
in the ranking distribution
of the intracellular concentrations.
This region is fairly described by a power-law distribution.
And, as observed in Fig. \ref{fig:concentration_evo}, 
it widens as the simulation proceeds.
Nevertheless, a well defined cutoff in the low concentration region is always present.
This is consistent with the results previously reported by Kaneko and Furusawa (2006) for a similar simulation \cite{furusawa06}.

\begin{figure}[!htbp]
\includegraphics[scale=1]{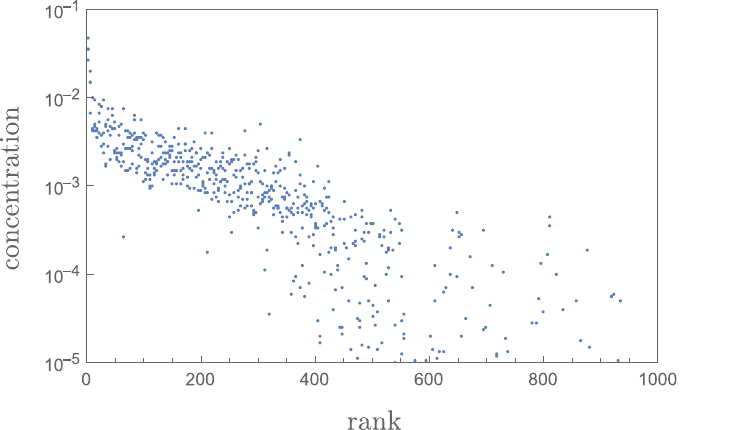}
\centering
\caption[Correlation between nodes degree and intracellular concentration.]{Correlation between nodes degree and intracellular concentration. Here the different species are ranked according to its node overall degree. These concentrations in particular correspond to a $1000$ nodes network at the generation $g=4000$.}
\label{fig:degcon_corr}
\end{figure}

\begin{figure}[!htbp]
\includegraphics[scale=1]{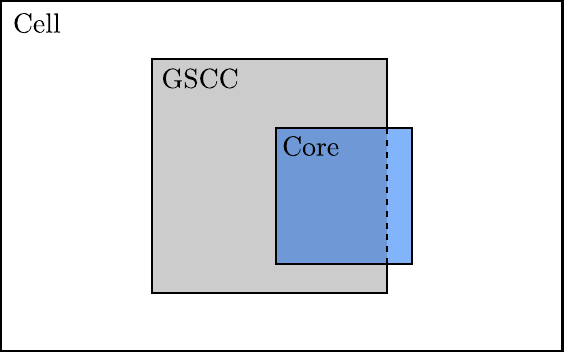}
\centering
\caption[Illustrative Venn diagram of the sets of nodes of the whole network, the giant strongly connected component and the core subgraph generated from the highest connected nodes.]
{
Illustrative Venn diagram showing the sets of nodes in the whole network
(white rectangle), 
the giant strongly connected component 
(gray rectangle)
and the core
(blue rectangle).
}
\label{fig:venn_diagram}
\end{figure}

\begin{figure}[!htbp]
\includegraphics[scale=1]{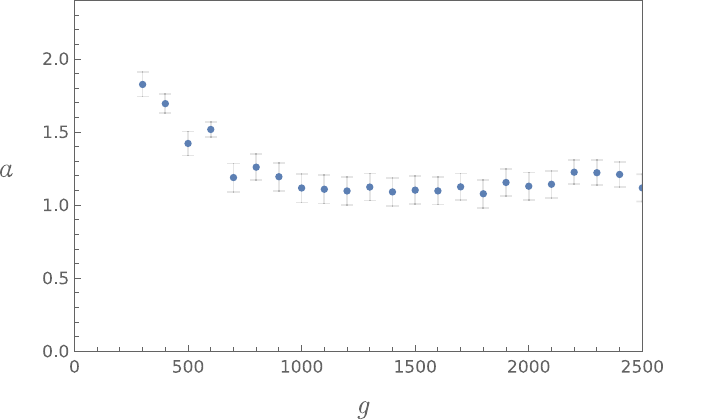}
\centering
\caption[Evolution of the exponent of the power-law distribution fit to the higher connectivity nodes.]
{Evolution of the exponent of the power-law distribution for the core. 
Gray bars denote the fit error. 
It is shown the result for a $500$ nodes network.}
\label{fig:alpha_evo}
\end{figure}

A deeper analysis 
revealed a positive correlation between concentrations and the overall node connectivity, as observed in Fig. 
\ref{fig:degcon_corr}.
This correlation is developed over the course of the simulation,
becoming distinct
for the set of nodes ranked with higher overall connectivity.
We have found that, 
in general, 
this set comprises between
20 and 30 percent of the graph nodes.
We labeled the corresponding
subgraph as \textit{core}.
It starts by being disconnected,
but becomes weakly connected around the $1.5N$ generation,
while the core complement remains disconnected 
the whole simulation.
The core is almost completely contained in the \textsc{gscc} of the whole network
as represented in Fig. \ref{fig:venn_diagram}.
The nodes that are in the core but not on the \textsc{gscc}
mostly correspond to nutrient species,
which by definition cannot be in a strongly connected component.
Following exclusively the behavior 
of the concentrations in the core,
it is observed that their ranked distribution
reaches a stable pure power-law with an exponent near $-1$ around the $2N$ generation
\footnote{The exponent is found using the Python library \texttt{plfit.py}, an implementation by Adam Ginsburg of the general algorithm presented by Clauset et al. (2009) \cite{clauset09}.}. 
This is illustrated in Fig. \ref{fig:alpha_evo}.

As we noted in Sec. \ref{sec:<d>},
the evolution 
of the whole network into a small-world structure
can be completely determined by the changes in the \textsc{gscc}.
Here we find that the main functional characteristics
of the cell are dictated by the nodes in the core.
Moreover,
the structure presented in Fig. \ref{fig:venn_diagram} 
resembles the connectivity structure 
of real metabolic networks
found in Ref. \cite{hong03}.

\section{Summary and discussion}
\label{sec:conclusions}
\begin{figure*}[!htbp]
\includegraphics[scale=1]{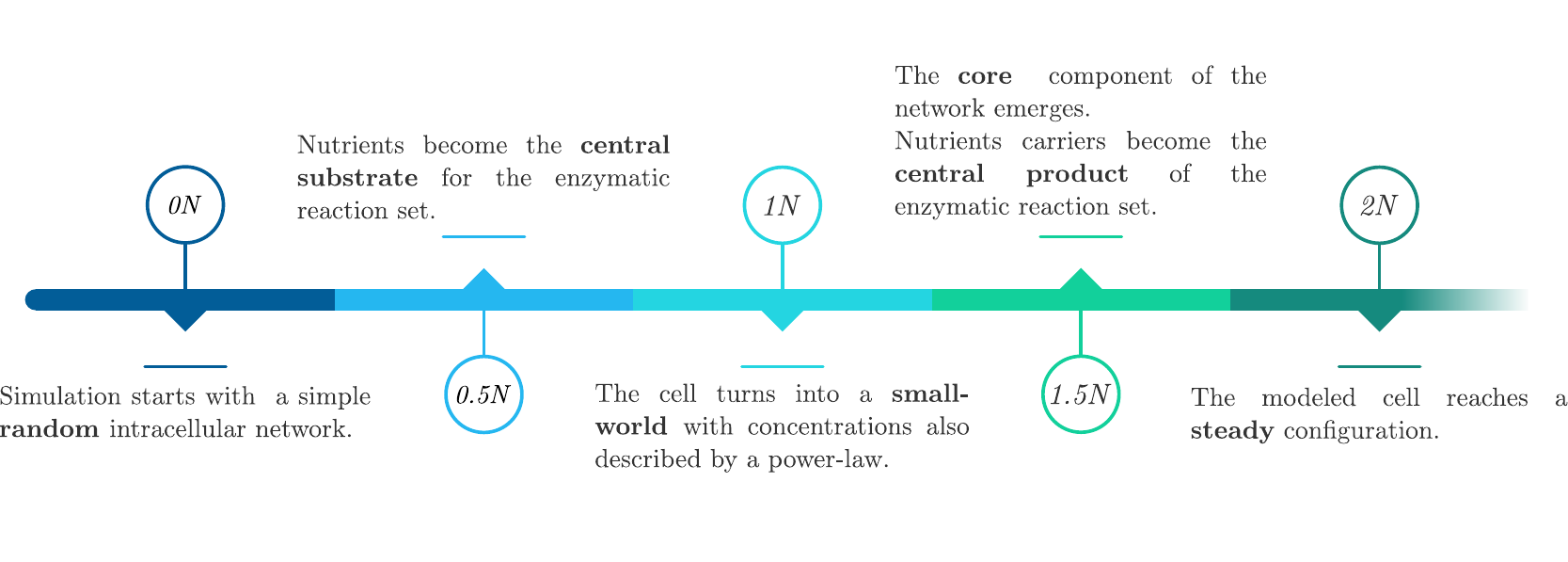}
\caption[Timeline of the modeled cell during the mutation-selection process.]
{Timeline of a modeled cell during the mutation-selection process.}
\label{fig:timeline}
\end{figure*}

The results presented in the previous section are summarized in the evolutionary timeline shown in Fig. \ref{fig:timeline}.
Let us delineate the process,
recalling first that 
we have found that
the number of generations 
at which the main phases of the evolution 
take place
can be parametrized by $N$,
the number of species in the modeled cells
(in particular $N$ taking here the values 250, 500 and 1000).

In the beginning,
i.e., at $0N$ generations,
cells were considered to be random networks. 
The idea of randomness as a precursor of life
can be traced back to the pioneering works of Kauffman,
whose theories centered on properties of 
random directed graphs \cite{Kauffman86}.
Several other mathematical models
have shown that a primitive metabolic system can arise from 
a stochastic configuration of inanimate elements \cite{steel04,filisetti14}.
Following this line,
in this work we have shown that
from a primordial cell,
modeled as a random set of catalytic reactions,
a self-reproducing system can arise
in a Darwinian process
of mutation and selection 
that favours cells with the highest replication rate.
As we will see,
we found 
a better correspondence with observational data
when the initial cell 
is constituted by a rather small number of catalytic reactions.
From an evolutionary perspective,
it implies
starting from simple cells
and evolving via mutation–selection dynamics into more complex ones.

After the simulation is set up,
the first distinct signal of organization
appears by the $0.5N$-th generation,
when nodes corresponding to nutrient species achieve the highest outward centrality among all nodes.
This means that, 
at this stage,
the mutation-selection mechanism
picks up
cells shortening metabolic pathways
that link nutrients to every other species.
Bearing in mind 
that in our simple model
only the species permeated from outside the cell 
(i.e., the nutrients)
act as precursors for the synthesis of the rest of the species,
our result is in agreement with the one found for the metabolic networks of living organisms, 
where the dominant `outward' central metabolites correspond mainly to building-block molecules.
For example,
in the case of the metabolic network of the \emph{Escherichia coli}, the top `outward' central metabolites correspond to pyruvate and acetyl-CoA, 
along with other intermediary metabolites \cite{hong03}.

Around the $1N$-th generation of the simulation,
the degree distribution of the cellular networks
gets described by a power-law 
(coming from an initial binomial distribution). 
This indicates the emergence of the scale-free structure 
universally associated with the so-called 
\emph{small worlds}.
Such connectivity is one of the main structural characteristic of real metabolic networks, 
and have been linked to the ability of the cell metabolism to rapidly react to both internal and external perturbations 
\cite{fell01,watts98}.
We have found that to obtain a pure power-law for the degree distribution
of a modeled cell,
as observed in real metabolic networks
\cite{broido18},
it is convenient to start the simulation
with a low degree of connectivity.
This is suitable for describing
the evolution of a cell
which starts as a set of elementary 
and mostly independent chemical reactions
and, 
while competing to be the fittest one,
becomes more complex with every passing generation,
exhibiting a richer structure of interconnected enzymatic reactions.
It is important to note
a key difference in the interpretation
of the scale-free property in our study
with respect to that discussed in Albert-Barabási works 
based on the preferential attachment model of scale-free networks
\cite{barabasi16}.
According to the latter,
it is hypothesized that in a metabolic network the highly connected metabolites should be the oldest phylogenetically,
which is well supported by experimental analysis \cite{fell01}.
Nevertheless,
in the catalytic reaction network model 
there is no temporal distinction among nodes,
thus,
the highly connected ones arise solely by their functional roles 
in the reaction network,
displaying a complementary mechanism
by which such hub nodes can develop.
With regard to this point,
it is also worth noting that
the emergence of the power-law for the degrees distribution
coincides 
with the rise 
of a scaling region
in the ranked concentrations distribution
described also by a power-law.
As mentioned, 
this kind of behavior
is an universal fingerprint of the intracellular chemical composition of living organisms \cite{jeong00,shumpei18}.

Near the $1.5N$-th generation,
the subgraph of the nodes with the highest connectivity degree
becomes connected, 
indicating the emergence of a tightly connected 
\emph{core} 
component in the network.
This is consistent with
the fact that, 
for many living cells,
there is a small number of keys metabolites
(a few dozens)
that exhibit the highest degree of connectivity
\cite{fell01,hong03}.
Interestingly,
we also found that,
at this stage, 
the nodes belonging to this core 
mostly correspond to those species exhibiting the highest concentrations.
The key functional role of the core in the modeled metabolism 
is highlighted by the fact that 
the subgraph which complements the core in the network
remains disconnected throughout the whole evolution.
In this regard,
it should be mentioned
that using exclusively a criterion based on the clustering distribution,
in our modeled cell we did not find
the structure of nested subnetworks
typical for the metabolism of living cells
\cite{ravasz02}.
Nevertheless,
this key difference 
has been observed before in similar processes 
where a network is gradually modified,
particularly,
in the Albert-Barabási preferential attachment model
\cite{barabasi16}.
Besides using another algorithms to search for hierarchical structure,
it motivates us to study modifications of the model to include,
for example,
cost of maintenance of metabolic pathways and
to simulate cellular adaptation to a given environment along with the cellular growth optimization. 
These mechanisms had been directly associated with the development of a hierarchical structure in bacterial metabolic networks \cite{goodman19}.
Back to the functional role of the core,
it is important to note that
this is the same evolutionary phase
when the nutrient carriers finally become central
as product of the chemical reactions.
This seems to differ
from what it is reported in Ref.\cite{hong03},
where it was found that
the top most `inward' central metabolites
in the \emph{Escherichia coli} metabolic network
correspond to 
the same intermediary metabolites
dominating the outward centrality.
Most of these metabolites
are part of the so-called `central metabolism', 
namely the glycolysis and tricarboxylic acid cycle pathways.
One reason for the apparent discrepancy between their and our results 
is that in our model
the `catabolic-like' reactions 
which could lead to this behavior, 
are deliberately excluded 
by restricting nutrients from being product of any reaction.
Recall that this constraint is enabled here
because we do not aim to recreate the inner cellular dynamics
on the whole,
but rather to focus solely on the traits 
leading to steady cell growth.
Nonetheless, 
this restriction in the model reveals the key role of the carrier species in the self-regulation mechanism of the cells 
during the mutation-selection process.
As a consequence of evolutionary pressure in favor of a faster growth rate,
mutations to the cells that promotes a more efficient nutrient consumption and transformation prevail.
In our simulations 
this is directly reflected 
in the fact
that the outward closeness centrality 
ended dominated by the nutrients species.
Still,
in order to maintain such an optimal growth rate,
the needed nutrient supply for the remaining species 
must be preserved
by an equally efficient nutrient inflow from outside the cell, 
which in this model is regulated solely by the carrier species.
We believe this is the reason why they end dominating the inward centrality.
Since the carriers are part of the core,
the fact that this happens concurrently with the emergence of the core
is unlikely to be a coincidence.

Within the constraints of our simulation, 
the final stage 
of the evolution of the modeled cell
is found to start around generation $2N$,
when the exponent in the power-law distribution 
describing the concentrations probability
for the core
reaches an stable value very close to $-1$.
This case,
corresponding to what is known as Zipf's law,
is special because is ubiquitous 
in systems where inner competition
lead to changes in their components size,
either by aggregation or fracture.
Such a simple scaling in the ranked probability distributions
has been observed in a broad range of domains like
word frequencies in a text, 
city and firm sizes,
incomes,
amino acids sequences,
neural activity,
hub traffic and social contacts networks.
This pervasive presence of the Zipf's law
is expected to be related with a fundamental
and universal mechanism for emergence of complexity 
\cite{cristelli12}.
For our simulation,
since the core component seems to dictate the performance of the whole cell,
the fact of settling down into this steady state 
implies that a configuration is reached
when new mutations not longer impact 
the reaction kinetics  
supported by the inner structure of our modeled cell.
This could be an instance of the
‘cost of complexity’ hypothesis
\cite{orr00}
at cellular level.

\section{Acknowledgments}

We thank Andrea Rodr\'iguez, Paulina Gonz\'alez, Roberto S\'aenz  and Sara Centeno 
for their useful discussions.
We also acknowledge HypernetLabs and Google Cloud Services for allowing us to use their computational facilities to run most of our simulations.
The work of D. R-O was supported by a CONACyT grant for graduate studies.

\bibliography{references}

\end{document}